\documentclass[aps,prx,article,twocolumn,showpacs,preprintnumbers,amsmath,amssymb,superscriptaddress]{revtex4}

\date{\today}
\usepackage{epsfig}
\usepackage{subfigure}
\usepackage{graphicx}
\usepackage{dcolumn}
\usepackage{bm}
\usepackage[colorlinks,linkcolor=blue,hyperindex,CJKbookmarks]{hyperref}
\usepackage{float}
\usepackage{hyperref}
\usepackage{comment}
\usepackage{color}
\newcommand{\tJ}{$t\small{-}J$}

\newcommand{\tJs}{$t\small{-}J\small{-} 3s$}

\newcommand{\Akw}{$A({\bf k},\omega)$}

\begin{document}
\title{Influence of Magnetism and Correlation \\ on the Spectral Properties of Doped Mott Insulators}
\author{Yao Wang}
\affiliation{Department of Applied Physics, Stanford University, California 94305, USA}
\affiliation{Stanford Institute for Materials and Energy Sciences, SLAC National Laboratory and Stanford University, Menlo Park, CA 94025, USA}
\affiliation{Department of Physics, Harvard University, Cambridge, Massachusetts 02138, USA}
\author{Brian Moritz}
\affiliation{Stanford Institute for Materials and Energy Sciences, SLAC National Laboratory and Stanford University, Menlo Park, CA 94025, USA}
\affiliation{Department of Physics and Astrophysics, University of North Dakota, Grand Forks, North Dakota 58202, USA}
\author{Cheng-Chien Chen}
\affiliation{Department of Physics, University of Alabama at Birmingham, Birmingham, Alabama 35294, USA}
\author{Thomas P. Devereaux}
\affiliation{Stanford Institute for Materials and Energy Sciences, SLAC National Laboratory and Stanford University, Menlo Park, CA 94025, USA}
\author{Krzysztof Wohlfeld}
\email[Corresponding author: krzysztof.wohlfeld@fuw.edu.pl]{}
\affiliation{Institute of Theoretical Physics, Faculty of Physics, University of Warsaw, Pasteura 5, PL-02093 Warsaw, Poland}

\date{\today}
\begin{abstract}
Unravelling the nature of doping-induced transition between a Mott insulator and a weakly correlated metal is crucial to 
understanding novel emergent phases in strongly correlated materials. For this purpose, we study the evolution 
of spectral properties upon doping Mott insulating states, by utilizing the cluster perturbation theory on the Hubbard and $t$--$J$-like models.
Specifically, a quasi-free dispersion crossing the Fermi level develops with small doping,
and it eventually evolves into the most dominant feature at high doping levels.
Although this dispersion is related to the free electron hopping,
our study shows that this spectral feature is in fact influenced inherently by both electron-electron correlation and 
spin exchange interaction: the correlation destroys coherence, while the coupling between spin and mobile charge restores it in the photoemission spectrum.
Due to the persistent impact of correlations and spin physics, the onset of gaps or the high-energy anomaly in the 
spectral functions can be expected in doped Mott insulators. 
\end{abstract}
\maketitle

\section{Introduction}
One central question in condensed matter physics is the origin of high-temperature superconductivity discovered 
about thirty years ago in copper oxides. Due to the combinatorially large degrees of freedom intrinsic in this 
quantum many-body problem, a microscopic first-principles study is impractical~\cite{RevModPhys.70.1039}.
On the other hand, it is believed that the underlying physics can be understood in terms of a minimal two-dimensional 
(2D) Hubbard model~\cite{lee2006doping}. In the limit of strong Hubbard repulsion, its low-energy physics can be 
further simplified into that of the \tJ\ model with a perturbative projection of double occupancies. Unlike the undoped 
(or half-filled) limit, the physics of collective excitations and emergent quasi-particles upon doping Mott insulators 
remains an open question for both models. Advancing the knowledge of how their spectral features evolve with 
doping is significant to understand intriguing emergent phenomena that can be found in strongly correlated materials.

\begin{figure*}
\begin{center}
\includegraphics[width=14cm]{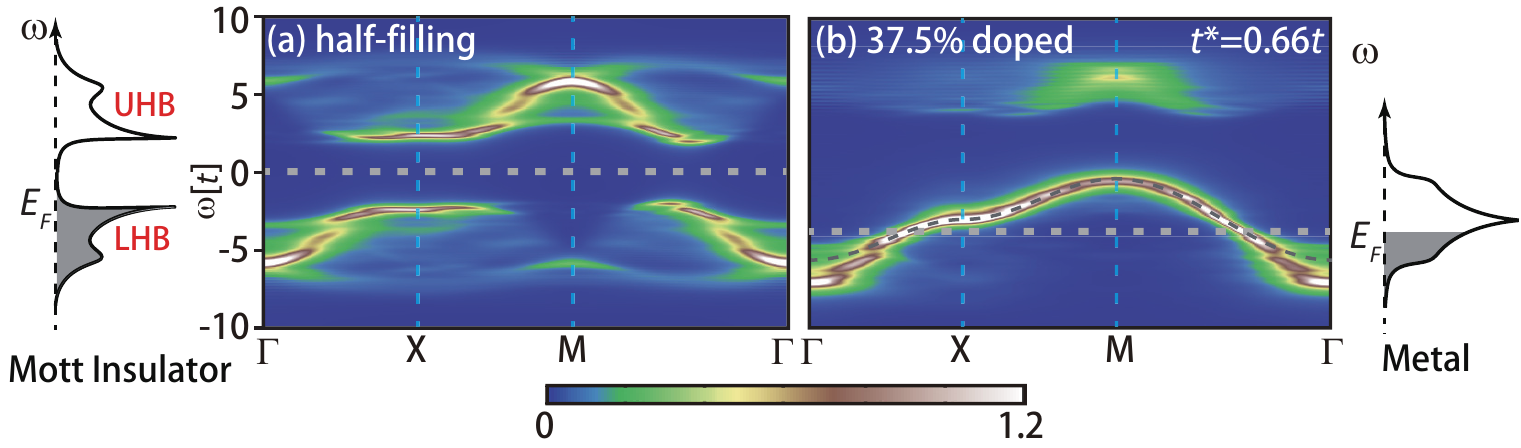}
\caption{\label{fig:1} Spectral function \Akw\ of the Hubbard model calculated by cluster perturbation theory (CPT) 
on a $4\times4$ square lattice at (a) half filling and (b) 37.5\% hole doping. The spectral features at half filling are 
dominated by the spin physics including the spin-polaron and 3-site term dispersions. The 37.5\% doped system 
shows a cosine-like quasi-free dispersion (the dashed gray line) described by a 2D nearest-neighbor tight-binding 
model with renormalized hopping $t^{*} = 0.66t$. The calculations adopt an on-site Hubbard interaction $U =8t$ 
and a Lorentzian broadening $\Gamma=0.15t$. The horizontal dashed lines denote the Fermi level. The insets sketch 
respectively the density of states for a Mott insulator and a metal.
}
\end{center}
\end{figure*}

In order to show the complexity of this problem in more detail, let us first concentrate on the spectral properties 
of the undoped limit of the Hubbard model [see Fig.~\ref{fig:1}(a)]. In this limit, the ground state of the Hubbard model 
is (Mott) insulating in the presence of strong on-site Coulomb repulsion.
The charge carriers are localized with
the valence and conduction bands separated by the so-called Mott gap [see Fig.~\ref{fig:1}(a)].
Moreover, this ground state exhibits long-range antiferromagnetism arising from strong correlation 
effects~\cite{Auerbach1994}. The elementary excitations (magnons) lie exclusively in the spin channel
and are of collective nature.
Consequently, the single-particle dynamics visible in the spectral function displays a dominance of spin physics.
This includes the lower-binding-energy spin-polaron~\cite{Schmitt:1988SCBAThry, Kane:1989SCBAThry, 
Martinez:1991SpinPolaron, Efstratios:2007SCBA, Zemljic:2008, Preuss:1995SDW, macridin2007high}, 
with the charge solely moving by coupling to magnons, 
and the higher-binding-energy intra-sublattice hopping mostly stemming from the so-called three-site 
terms~\cite{stephan1992single, tJ3s, Bala:1995ttprJSCBA, wang2015origin}, see Fig.~\ref{fig:1}(a).

\begin{figure*}[ht!]
\begin{center}
\includegraphics[width=16cm]{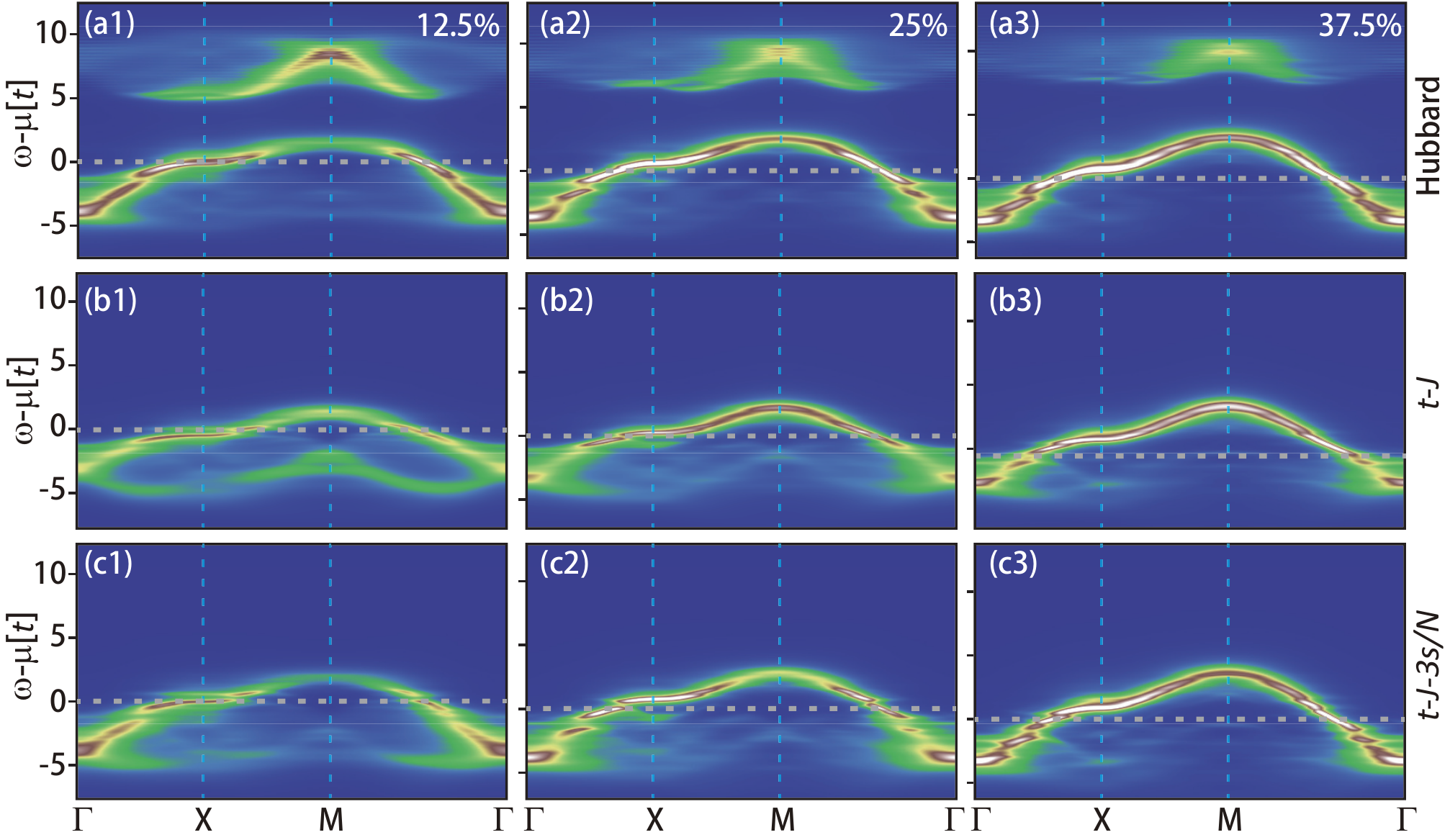}
\caption{\label{fig:2} Spectral function \Akw\ calculated 
by CPT at 12.5\% (left panels), 25\% (middle panels) and 37.5\% (right panels) hole doping for (a1-a3) the Hubbard model, (b1-b3) the \tJ\ model, and (c1-c3) the normalized \tJs\ model.
Spin exchange $J=4t^2/U = 0.5t$ is adopted for the \tJ\ and \tJs\ models.
The horizontal dashed lines denote the Fermi level $E_F$.}
\end{center}
\end{figure*}

This situation changes drastically when additional charge carriers are introduced: the long-range antiferromagnetic order 
diminishes at a few percent doping, and other competing phases such as stripe/charge order
or $d$-wave superconductivity could emerge~\cite{lee2006doping}. While simulating these broken symmetry states 
in the thermodynamic limit is a challenge for advanced numerical calculations~\cite{vojta2009lattice}, 
on a finite cluster a ground state lacking a broken symmetry can be realized~\cite{LeBlanc2015}. 
It turns out that the Hubbard spectral function in this case 
appears at first sight to be relatively simple: 
except for the spectral weight located well below the Fermi level and close to the $\Gamma$ point,
the dominant spectral feature below the Mott gap follows a cosine-like dispersion as that of a tight-binding model with a renormalized bandwidth. 
This {\it quasi-free dispersion} is well-visible already at 12.5\% doping and it absolutely dominates the spectrum 
at 37.5\% doping [see Fig.~\ref{fig:1}(b) and Fig.~\ref{fig:2}(a)]; this dispersion also seems to be the sole feature that crosses the Fermi level in all of the above spectra.

As suggested by the above results, while the spin physics is crucially important for understanding the undoped spectrum, 
it does not seem to play a similar dominant role in understanding the doped spectrum -- at least for doping levels of 
12.5\% or higher. Such a result stays in stark contrast with several other recent numerical simulations
of the Hubbard model. These finite-size calculations explore the evolution of the spin response upon hole or electron doping,
and the results indicate the persistence of collective spin excitations until about 40\% hole-doping~\cite{jia2014persistent, Kung2015,
huang2016numerical, TKLee2016, Kung2017}.
Moreover, a large number of resonant inelastic x-ray scattering (RIXS) experiments also have revealed the persistence of 
a ``paramagnon'' dispersion in some areas of the Brillouin zone upon both hole- and electron-doping the cuprates~\cite{Braicovich2010, le2011intense,
dean2013persistence, Dean_PRL_2013, lee2014asymmetry, Guarise2014, ishii2014high, Huang2016, Minola2017}. 
Another experimental evidence showing the existence of strong spin fluctuations is the widely observed ``hourglass'' structure in inelastic neutron scattering at 1/8 doping~\cite{tranquada2004quantum, tranquada2007neutron}. 
All these observations bring us to the two main questions of the paper: What is the nature of the dominant quasi-free dispersion feature in the 
doped Hubbard spectral function? Could it be intrinsically influenced by the spin physics?

In this work, we intend to answer the above questions by investigating in detail the origin and evolution of the quasi-free dispersion 
feature upon doping the Hubbard model. The focused regime of our study contains the optimal doping at about 12.5\% 
and extends up to about 37.5\% doping, above which the collective spin excitations are no longer well-visible in the 
Hubbard model. Naturally, related studies have been partially performed already in the 
1990s~\cite{Stephan1991, Moreo1995, Haas1995, Wen1996, Duffy1997} and around 
2000s~\cite{lee2006doping, Byczuk2007, Moritz:2009fb}, and also in a few very recent 
contributions~\cite{kohno2015spectral, Sakai2016L, Sakai2016, shastry2017}.
Nevertheless, we believe that recent progress in numerical techniques combined with a detailed analysis of the 
obtained results can give new insight into this problem. In particular, the recent success in calculating the spectral 
functions of the $t$--$J$ model using cluster perturbation theory (CPT)~\cite{wang2015origin, kohno2015spectral}, 
the method designed for studying the Hubbard spectral function~\cite{Senechal:2000fg, Maska1998, Senechal:2002fr, 
aichhorn2005weak, senechal2012book, pairault1998strong, Kohno:2012PRL, PhysRevB.90.035111, wang2015origin}, enables a reliable comparison 
between the Hubbard and $t$--$J$ models at the same stage. 
Since understanding the $t$--$J$ spectra is far simpler than in the case of the Hubbard, an intuitive explanation of 
the various spectral features observed in the Hubbard model is then made possible. 

The rest of the paper is organized as follows.
In Sec.~II we give a brief overview of the Hubbard model and the CPT method.
In Sec.~III we present the numerical results showing the onset of a quasi-free dispersion upon doping 
the Hubbard model and compare its spectral functions to those of the $t$--$J$-like models.
In Sec.~IV we provide an interpretation of the $t$--$J$ model spectra, focusing especially on the nature 
of the quasi-free dispersion. Finally, we conclude the paper in Sec.~V by summarizing our main results. 

\section{Hubbard Model and Cluster Perturbation Theory}
To describe the strongly correlated nature of doped Mott insulators, four-fermion interactions have to be considered 
on top of the tight-binding terms. For this purpose, the 2D Hubbard Hamiltonian is usually used as a 
minimal model~\cite{anderson1987resonating, Zhang:1988jf, Eskes:1988ef}:
\begin{equation} \label{eq:hubbard}
\mathcal{H}=-t\! \! \sum_{\langle{\bf i},{\bf j}\rangle,\sigma} \!\left(\!c^\dagger_{{\bf j} \sigma} c_{{\bf i} \sigma} \!+\!h.c.\! \right) \!+ U\sum_{{\bf i}}\left(\!n_{{\bf i}\uparrow}\!-\!\frac12\!\right)\left(\!n_{{\bf i}\downarrow}\!-\!\frac12\!\right),
\end{equation}
where $c^\dagger_{{\bf i}\sigma}$ ($c_{{\bf i}\sigma}$) denotes the creation (annihilation) operator of 
spin $\sigma$ at site $\textbf{i}$, and $n_{{\bf i}\sigma} \equiv c^\dagger_{{\bf i}\sigma}c_{{\bf i}\sigma}$ 
is the corresponding density operator; $t_{\bf ij}$ is the hopping amplitude, and $U$ is the on-site Hubbard 
repulsion strength. For the purpose of simplifying the degrees of freedom and comparing with \tJ-like spin models, 
here we consider only nearest-neighbor hopping $t_{\bf \langle ij\rangle}\!\equiv\!t$ 
and neglect longer-range terms such as $t'$. Thus, the carriers are equivalent under a particle-hole 
transformation. Therefore, the specific difference between electron- and hole-doped Mott insulators 
is not discussed in the scope of this work. Furthermore, finite $t'$ merely changes the uncorrelated 
physics of the Hubbard model, so its effect on the spectral features is relatively well-understood in our scope.

The spectral function \Akw\ of the Hubbard model has been calculated by various numerical methods, 
such as exact diagonalization (ED)~\cite{PhysRevB.44.10256, Dagotto1992, RevModPhys.70.1039}, 
quantum Monte Carlo~\cite{PhysRevB.47.1160, Preuss:1995SDW, Groeber2000, Moritz:2009fb, becca_sorella_2017}, 
dynamical mean-field theory~\cite{Georges1996, kancharla2008anomalous, weber2010strength, sakai2009evolution, sakai2010doped}, 
CPT~\cite{Senechal:2000fg,Senechal:2002fr,aichhorn2005weak, senechal2012book,pairault1998strong, Kohno:2012PRL, PhysRevB.90.035111}, and several 
others~\cite{Schollwoeck2005, yuan2005doping,arrigoni2009phase, gull2013superconductivity, han2016charge}. For the purpose of this paper, we believe CPT is the most suitable method, 
as it can produce zero-temperature spectra with continuous momentum resolution. This fine energy-momentum structure 
then allows a detailed characterization of various features and mechanisms.

The CPT method usually proceeds with solving the model Hamiltonian by ED on finite-size clusters with open-boundary 
conditions~\cite{Potthoff_PRB_2003}. An approximate infinite-size lattice Green's function is constructed using 
the small-cluster results by treating inter-cluster hopping with a strong-coupling perturbation 
theory~\cite{pairault1998strong, pairault2000strong}. The zero-temperature spectral function \Akw\ can then be 
obtained accordingly~\cite{Senechal:2000fg, Senechal:2002fr}. This method would yield numerically exact results 
in both the noninteracting ($U=0$) and strong-coupling ($t=0$) limits. When $U$ and $t$ are both finite, short-range 
correlations caused by strong interaction are incorporated in finite-size clusters, and long-distance effects are accounted 
for by perturbation theory, together rendering the method adequate for intermediate coupling.
Although CPT was designed for models with local interactions, it has been shown that this method also can correctly capture 
the spectral features of \tJ\ models with the nearest-neighbor hopping and interactions~\cite{wang2015origin}. 
Here we apply CPT to compute the spectral functions based on $4\times4$ clusters.

\begin{figure*}[!ht]
\begin{center}
\includegraphics[width=16cm]{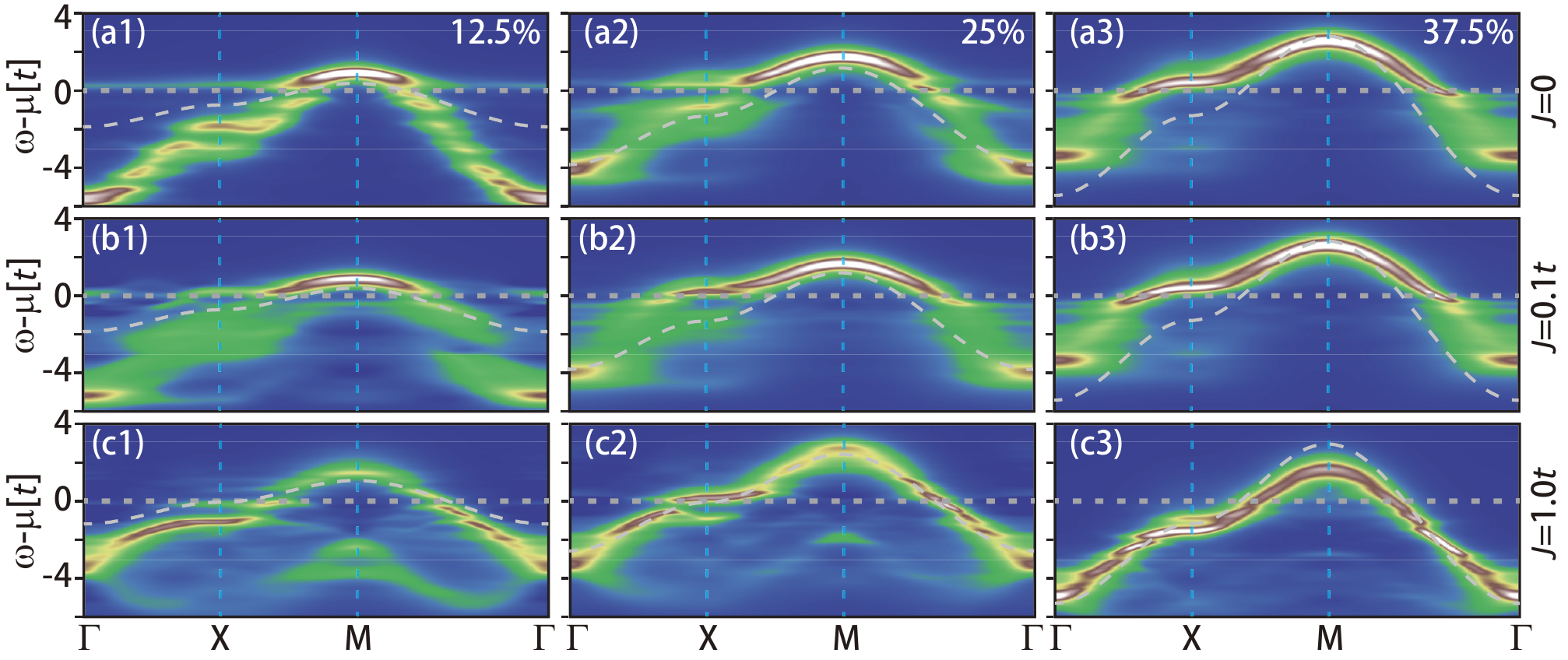}
\caption{\label{fig:3}
Spectral function \Akw\ calculated by CPT for the \tJ\ model at 12.5\% (left panels), 25\% (middle panels) and 37.5\% (right panels) hole dopings with (a1-a3) $J=0t$, (b1-b3) $J=0.1t$, and (c1-c3) $J=t$. 
The dashed curves show the 2D nearest-neighbor tight-binding dispersions with a renormalized hopping $t^{*} = g_t t$, where $g_t$ is the doping-dependent Gutzwiller factor in the Gutzwiller mean-field approximation~\cite{Zhang1988}.
The horizontal dashed lines denote the Fermi level.
}
\end{center}
\end{figure*}

\section{Hubbard spectral function and its comparison with the $t$--$J$ spectrum}
Figures~\ref{fig:2}(a1)-(a3) show the spectral functions calculated by CPT for the Hubbard model with $U=8t$ at three hole dopings: 12.5\%, 25\%, and 37.5\%.
Comparing Fig.~\ref{fig:2}(a1) with Fig.~\ref{fig:1}(a), we first note that three features of the undoped Mott insulator are also visible in the doped system. These include the relatively broad and 
largely incoherent upper Hubbard band, and the two features associated with the spin physics at half filling~\cite{wang2015origin}.
The latter features are solely seen in the high-binding energy part of \Akw\: mostly around the $\Gamma$ point, 
with much weaker contributions around the X / M points, see Fig.~\ref{fig:2}(a). 
Since these features are located far away from the Fermi level and, except close to the $\Gamma$ point, their spectral weight never dominates and decreases with increasing doping, we leave the understanding of their doping evolution for future study.

On the other hand, it turns out that already at 12.5\% doping the most apparent spectral feature is the one that can be associated with the so-called quasi-free dispersion defined in Sec.~I in the context of the spectrum calculated at 37.5\% doping [see also Fig.~\ref{fig:1}(b)].
This is because already in this case, except for a close vicinity to the $\Gamma$ point, 
the dominant spectral weight below the Mott gap is concentrated around the feature that qualitatively shows the same (cosine-like) momentum-dependence 
as predicted by the noninteracting (tight-binding) model. 
Naturally, there are also important differences between such a quasi-free dispersion feature and the bare tight-binding model:
the quasi-free dispersion shows not only a strong bandwidth renormalization, but also a large intrinsic spectral weight broadening.
In the following section we will investigate some of the physics behind this onset of the effective mass and finite lifetime of the electrons in the Hubbard spectral function. 

However, before discussing in detail the nature of the quasi-free dispersion, let us first consider the framework to simplify the problem. 
The lowest-order $t/U$ expansion of the Hubbard model leads to the so-called \tJs\ model with the Hamiltonian given by $\mathcal{H}_{t\!-\!J\!-\!3s}= \mathcal{H}_{t \!-\!J}+\mathcal{H}_{3s}$~\cite{chao1977tJmodel, chao1978canonical, belinicher1994consistent, belinicher1994range, stephan1992single, tJ3s, Bala:1995ttprJSCBA, Spalek1988, Szczepanski1990, Eskes1994, psaltakis1992optical}:
\begin{align}\label{tJH}
\mathcal{H}_{t\!-\!J}&=-t\!\sum_{\langle {\bf i},{\bf j} \rangle,\sigma}\! \left( \tilde{c}^\dagger_{{\bf j} \sigma} \tilde{c}_{{\bf i}\sigma}\! +\!h.c.\right)\!+\!J\sum_{\langle {\bf i},{\bf j}\rangle}\left( \textbf{S}_{\bf i} \! \cdot \! \textbf{S}_{\bf j}-\frac{\tilde{n}_i \tilde{n}_j}{4}\right)\!, \nonumber\\
\mathcal{H}_{3s}&= - \frac{J}{4}\!\!\sum_{\langle {\bf i},{\bf j}\rangle,\langle {\bf i},{\bf j}^\prime\rangle\atop {\bf j}\neq {\bf j}^\prime,\sigma}\! \! \left(\tilde{c}^\dagger_{{\bf j}^\prime\sigma}\tilde{n}_{{\bf i}\bar{\sigma}}\tilde{c}_{{\bf j}\sigma}\! - \! \tilde{c}^\dagger_{{\bf j}^\prime\sigma} \tilde{c}^\dagger_{{\bf i}\bar{\sigma}} \tilde{c}_{{\bf i}\sigma} \tilde{c}_{{\bf j}\bar{\sigma}}\right)\!,
\end{align}
where $\textbf{S}_{\bf i}\cdot\textbf{S}_{\bf j}\! =\! S_{\bf i}^zS^z_{\bf j}\!+\!\frac12\left(S_{\bf i}^+S_{\bf j}^-+S_{\bf i}^-S_{\bf j}^+\right)$, with $S_{\bf i}^z\!=\!(n_{{\bf i} \uparrow}\!-\!n_{{\bf i}\downarrow})/2$ and $S_{\bf i}^+\!=\!(S_{\bf i}^-)^\dagger\!=\!\tilde{c}^\dagger_{{\bf i}\uparrow} \tilde{c}_{{\bf i}\downarrow}$.
The constrained fermionic operators acting in the Hilbert space without double occupancies are defined as $\tilde{c}^{\dag}_{{\bf i}\sigma} = {c}_{{\bf i}\sigma}^{\dag}(1-n_{{\bf i} \bar{\sigma}})$. 

It is generally believed that the \tJs\ model describes most of the low-energy excitations in the Hubbard model, with the contribution of the upper Hubbard band being projected out. 
To justify this argument further, we compare the spectral functions of the Hubbard, the \tJ, and the renormalized \tJs\ models in Fig.~\ref{fig:2}. In panels (c1-c3) the spectral weight of the \tJs\ model is renormalized by the electron occupation $n_\textbf{k}$ of the Hubbard model.
Similar to the undoped case discussed in Ref.~\onlinecite{wang2015origin}, the renormalized \tJs\ spectral function displays both qualitative and quantitative agreements with that of the Hubbard model at the low binding energy ($\lesssim 4t$).
The only significant difference between these two spectra is the presence of the upper Hubbard band well above the Fermi level -- this part of the spectrum by construction cannot be captured by the \tJs\ model where doubly occupied states are integrated out.

\begin{figure*}[ht!]
\begin{center}
\includegraphics[width=14cm]{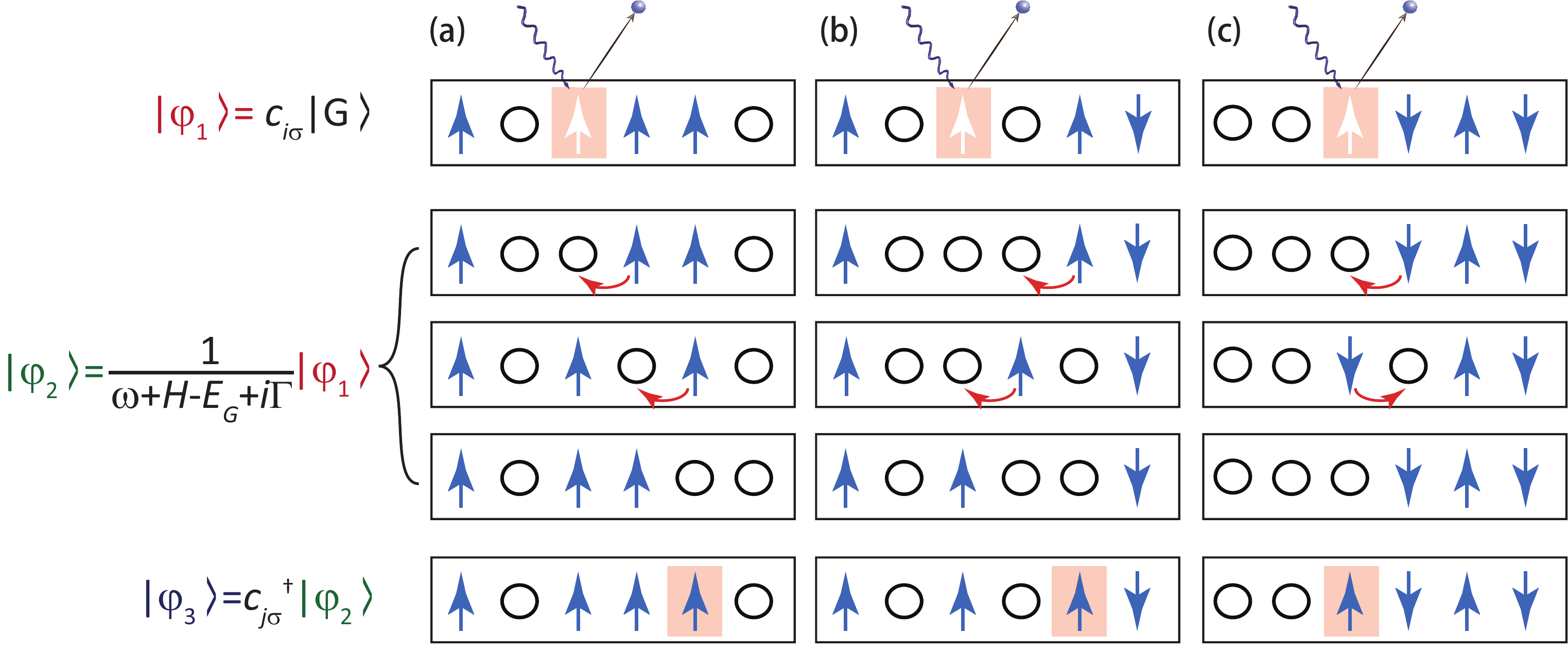}
\caption{\label{fig:4} Cartoons illustrating the motion of a photo-doped hole in the $t$--$J$ model with $J=0$ and finite density of holes in the ground state: 
(a) the coherent motion of a photo-doped hole allowed due to the existence of ferromagnetic clusters in the ground state $|G\rangle$ of a hole-doped $t$--$J$ model;
(b) the coherent motion of a photo-doped hole allowed due to the existence of empty sites in the ground state $|G\rangle$ of a hole-doped $t$--$J$ model;
(c) {\it incoherent} motion of a photo-doped hole along the retraceable paths (here of one lattice spacing). 
The photoemission spectrum is related to the imaginary part of the Green's function $\langle \varphi_3|\varphi_1\rangle$.
For simplicity, the cartoons only show hole motion along a 1D path in a 2D lattice.
}
\end{center}
\end{figure*}

More strikingly, the spectral features of the \tJ\ and the Hubbard models below the Mott gap still match qualitatively, despite of different intrinsic spectral weights due to distinct sum rules and the omission of the three-site terms in the former model. This agreement indicates that the nature of the quasi-free dispersion is also well-captured by the \tJ\ model upon doping.
Moreover, in both models the features associated with the spin physics (see discussion above) occur at relatively similar energies and momenta,
indicating that, unlike for the undoped case, the 3-site terms do not play an important role already for the 12.5\% doping.

\section{Understanding the $t$--$J$ model spectrum}
With the agreement between the main spectral features of the Hubbard and the \tJ\ models, we now turn to the understanding of the latter model with fewer degrees of freedom.
For this purpose, we calculate the spectral function for three different values of spin exchange $J/t=0$, $0.1$, and $1.0$, at three distinct hole-doping levels as shown in Fig.~\ref{fig:3} [the $J=0.5t$ spectra have been shown in Figs.~\ref{fig:2}(b1)-(b3)].
The spin exchange $J$ strongly modifies the shape of the spectral function for all doping levels,
accounting {\it inter alia} for the bandwidth renormalization of the quasi-free dispersion discussed earlier.
Moreover, even at the $J=0$ limit (below also called the ``constrained fermion model'',~cf.~Refs~\cite{Brinkman1970, Ruckenstein1988, Fang1989, Gros1989, Gehlhoff1995, Gurin2001, Foussats2002, Bejas2006}), the spectrum is different from the non-interacting fermionic model, see Figs.~\ref{fig:3}(a1)-(a3).

In the following subsection we concentrate on explaining the peculiar spectrum obtained in the $J=0$ limit, while in the next subsection we discuss the impact of a finite spin exchange $J$.
In order to streamline the discussion, we mainly focus on the ``most intensive feature'' found in the spectral functions which, except for the vicinity of the $\Gamma$ point, means looking at the quasi-free dispersion.

\subsection{Spectral Properties of the $t$--$J$ Model with $J=0$: Significance of Correlations}

To understand the spectral function of a doped system, we focus on the lowest canonical doping of 12.5\% carriers.
In this case, the spectrum of the \tJ\ model with $J=0$ is largely incoherent, where the most intensive peaks do not tend to form a continuous dispersion relation.
Below we refer to this feature as ``semi-coherence''~\cite{Footnote3} -- which denotes the case when the hole cannot move as a coherent quasiparticle in momentum space, 
irrespective of the existence of a well-defined quasiparticle at a particular momentum.
Moreover, there is a notable difference between the particle removal and addition spectra: the latter seems to be dominated by a single coherent band, 
although a striking dispersionless band can be seen just above the Fermi level. 
We note in passing that the latter mechanism may be the reason for the onset of a gap at the Fermi level in the Hubbard spectral function.

Before investigating in detail the origin of the ``semi-coherent'' spectrum, let us note that such a discontinuous dispersion of the most intensive feature can be fitted with a renormalized free band $\varepsilon_{\bf k} = - 2 t^{*} (\cos k_x + \cos k_y)$ with $t^{*} / t \simeq 0.83$. The renormalization is thereby relatively weak, in contrast to the factor $g_t \equiv t^{*} / t \sim 0.28$ predicted by a simple Gutzwiller mean-field picture at the same doping~\cite{Zhang1988}. 
This mismatch can be attributed to the invalidity of the Gutzwiller picture outside the vicinity of the Fermi level. 

At first sight, the onset of a ``semi-coherent'' spectrum can hardly be explained by existing theories.
On one hand, since the doped holes are no longer coupled to spin excitations when $J=0$,
one could ostensibly argue that only the hopping term $t$ of the \tJ\ model is relevant: 
even though the hopping of constrained fermions is allowed only when double occupancies are excluded,
this constraint might not affect hole propagation in a doped system.
Naively, the hopping of doped holes in the constrained fermion model could be equivalent to spinless fermions, which support free motion of doped holes.
However, such a simple picture would suggest just one unrenormalized free band in the spectral function, which is clearly not the case here.

On the other hand, we can recall the result from the paper by Brinkman and Rice~\cite{Brinkman1970},
where a single hole doped into the half-filled ground state of the constrained fermion model is considered.
When the doped hole moves around, it ``scrambles'' the spin pattern of one of the (degenerate) undoped ground states.
While naively such ``scrambling'' should be irrelevant when $J=0$ in terms of energy, one should note that for a coherent hole motion to take place, the spin pattern of initial and final state must be identical.
That is possible only when the ground state is ferromagnetic, or if the so-called Trugman paths are invoked~\cite{TrugmanPath}.
Both cases have very small contributions to the spectral function of the undoped constrained fermion model~\cite{Brinkman1970, Vojta1998, TrugmanPath}.
Thus, unless the undoped ground state is a fully polarized ferromagnet,
the spectrum is completely incoherent and does not support a quasiparticle solution.
However, this is not the situation we encounter, either. 

Instead, the situation for the \tJ\ model with $J=0$ and {\it finite} doping lies in between the above two extremes.
More precisely, we propose that the following three processes be crucial in explaining the spectral function of the doped constrained fermion model [see Fig.~\ref{fig:4}]:

Firstly, we have verified that the 12.5\% hole-doped ground state has significant ferromagnetic correlations between nearest neighbors~\cite{Footnote4}.
Such strong correlations are absent in the undoped case and can be understood as a consequence of the Nagaoka theorem: a single hole doped into the constrained fermion model on an infinite lattice fully polarizes the ground state. 
Due to the existence of finite ferromagnetic clusters in the ground state, a coherent motion of the photoinduced hole or electron becomes possible [see Fig.~\ref{fig:4}(a)], which (unlike the single hole case) has a significant contribution to the spectral function.

\begin{figure*}[th!]
\begin{center}
\includegraphics[width=12cm]{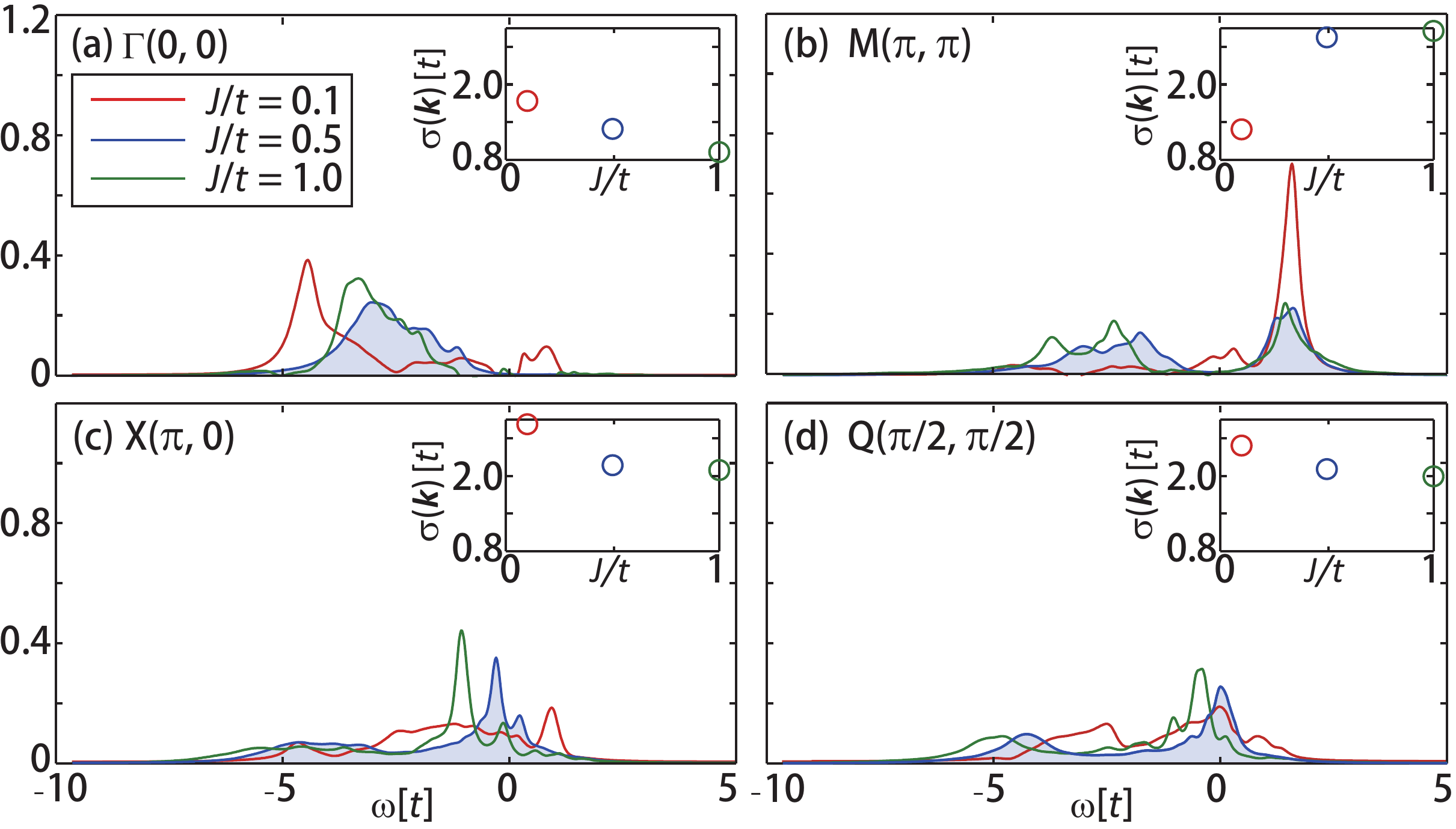}
\caption{\label{fig:EDC} Energy-dependent cuts (EDCs) of the \tJ\ model spectral functions at the 12.5\% doping level and for three distinct values of spin exchange $J$ at four distinct momenta ${\bf k}$: (a) $\Gamma$, (b) M, (c) X, and (d) Q. The insets show the variance $\sigma(\textbf{k})$ of the most intensive peak at each momentum calculated as a function of $J$ with $\Delta \omega = 4t$. The shaded blue curves denote the canonical $J=0.5$ scenario.}
\end{center}
\end{figure*}

Secondly, the empty sites in a doped ground state further support the coherent motion of charge carriers.
As shown in Fig.~\ref{fig:4}(b), these empty sites can mitigate the ``scrambling' 'of the spin pattern by the moving photo-doped holes or electrons.
Beyond the simple case of a 1D chain, such ``unscrambling'' process is expected to be more efficient in the 2D case, as a result of more possible paths.
This mechanism not only produces coherent motion but also reveals the asymmetry upon particle addition and removal: 
as long as the empty sites of the hole-doped state are within the 3rd nearest neighbors, the photoinduced electron can move without being hindered.
In terms of energy, the coherence is determined by the density of different states in a small range of energy. Therefore, the particle addition, in fact, reduces the number of configurations, and thus increases 
the coherence in contrast to a particle removal. This asymmetry is consistent with the spectral function, where the most intensive feature above the Fermi level is more coherent than that below the Fermi level.

Finally and perhaps most importantly, the photo-doped hole or electron also can move in an incoherent way, somewhat similar to the undoped case discussed in Ref.~\onlinecite{Brinkman1970}.
As shown in Fig.~\ref{fig:4}(c), for example, a photoinduced hole can move along the retraceable paths:
it first moves around by ``scrambling'' the spin pattern of the ground state, but then it heals this pattern by moving back along the same path to its original position.
Realizing that the moving hole couples to spin excitations (costing zero energy in the limit of $J=0$) at each step of its motion, we can understand that such a process should be incoherent.
While this incoherent process is not encountered for either spinless or spinfull free fermions, it has an important contribution to the total photoemission spectral weight of a doped constrained fermion model.
As the retraceable paths are dominant here, the bandwidth of such an incoherent spectrum in 2D should be reduced by a factor $\propto \sqrt{3} /  2 \simeq 0.87 $~\cite{Brinkman1970}.
Such a reduction plays dominant role in the observed band renormalization of $t^{*} / t \simeq 0.83 $. 

\subsection{Spectral Properties of the \tJ\ Model: Significance of Spin Physics}

We next discuss how the spectral function of the $t$--$J$ model changes once the spin exchange $J$ is finite and then increased from the realistic value of $J=0.5t$ to $J=1t$, see Figs.~\ref{fig:2}(b1)-(b3) and~\ref{fig:3}.
As before, below we discuss the lowest canonical doping of 12.5\% holes and mainly focus on the most intensive feature in the spectrum.

\begin{figure}[b!]
\begin{center}
\includegraphics[width=6cm]{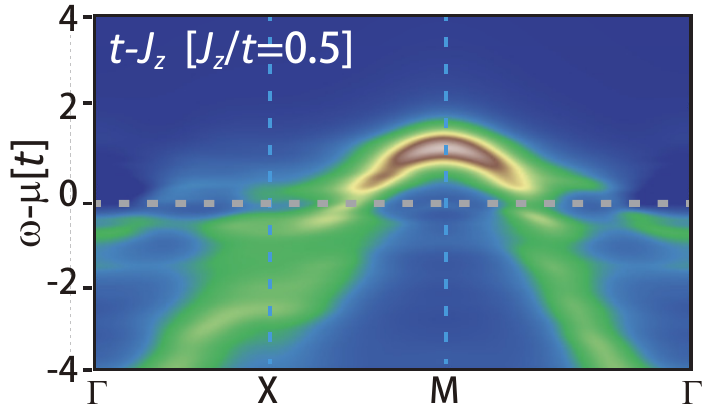}
\caption{\label{fig:6} Spectral function \Akw\ calculated by CPT for the $t$--$J^z$ model ({\it i.e.} the $t$--$J$ model without quantum spin fluctuations) with $J^z=0.5t$ at 12.5\% hole doping. The horizontal dashed line denotes the Fermi level.}
\end{center}
\end{figure}
It turns out that the potential quasiparticle character of the most intensive feature strongly depends on the value of $J$.
While the spectrum at $J=0$ exhibits a rather peculiar shape, denoted above as ``semi-coherence'', such a type of the spectrum is basically lost for a finite spin exchange energy [see Fig.~\ref{fig:3}].
Moreover, the widths of the most intensive peaks below the Fermi level in the spectral function are reduced with increasing $J$,
as seen in the energy-dependent cuts (EDCs) at selected high-symmetry momenta shown in Fig.~\ref{fig:EDC}.
To examine such an impact of spin exchange, we calculate the variance of the most intensive peak for each $\textbf{k}$ defined as
\begin{align}
\sigma(\textbf{k}) = \sqrt{\frac{\int_{\omega_0(\textbf{k})-\Delta \omega}^{\omega_0(\textbf{k})+\Delta \omega} 
A(\textbf{k},\omega) [\omega-\omega_0(\textbf{k})]^2 d\omega}{\int_{\omega_0(\textbf{k})-\Delta \omega}^{\omega_0(\textbf{k})+\Delta \omega} A(\textbf{k},\omega) \,d\omega} },
\end{align}
where $\Delta \omega = 4t $ and $\omega_0 $ is the peak position at each momentum. Here $\sigma(\textbf{k})$ gives a quantitative description of the coherence. 
The insets in Fig.~\ref{fig:EDC} show the variances calculated for three different values of $J$ at selected high-symmetry momenta.
As the spin exchange $J$ grows, it clearly reflects a trend of an increased quasiparticle coherence [equivalently a decreased $\sigma(\textbf{k})$] in the spectral function below the Fermi level. The opposite situation happens for the spectral functions above the Fermi level. Thus, the finite spin exchange tends to balance the coherence of the electron and hole spectrum. 

This importance of quantum spin fluctuations can be further confirmed by comparing the $t$--$J$ and $t$--$J^z$ model. 
The latter is achieved by turning off the quantum spin fluctuations $S_{\bf i}^+S_{\bf j}^-+S_{\bf i}^-S_{\bf j}^+$ in Eq.~\ref{tJH}.
Fig.~\ref{fig:6} shows the spectral function calculated in the $t$--$J^z$ model with the canonical $J^z=0.5t$ at the same doping level. Comparing Figs.~\ref{fig:2} and \ref{fig:3}, the variances of the spectra below and above the Fermi level are similar to that of a $t$--$J$ model with a much smaller $J$, indicating that the existence of quantum fluctuations is crucial for the onset of a more coherent quasiparticle in the photoemission spectrum. 

We also note that although the shape of the dispersion relation of the most intensive feature does not seem to heavily depend on $J$, its magnitude (the ``bandwidth'') is substantially enhanced with decreasing spin exchange $J$. 
This is especially visible when looking at the spectra at the M, $\Gamma$, and X points. For the latter two high symmetry points in the Brillouin zone the absolute value of the energy of the most intensive 
feature strongly increases once $J \rightarrow 0$ (e.g. at ${\bf k} = \Gamma$ the most intensive spectral feature is located at around $-3t$ for $J=0.5t$ and at almost $-6t$ for $J=0t$). 
At the same time, the energy of the most intensive feature at the M point only weakly reduces with decreasing $J$. This change in the magnitude of the dispersion relation as a function of $J$ is responsible for the observed
strong reduction of the Fermi surface for $J=0.5t$. 

\begin{figure}[t!]
\begin{center}
\includegraphics[width=6cm]{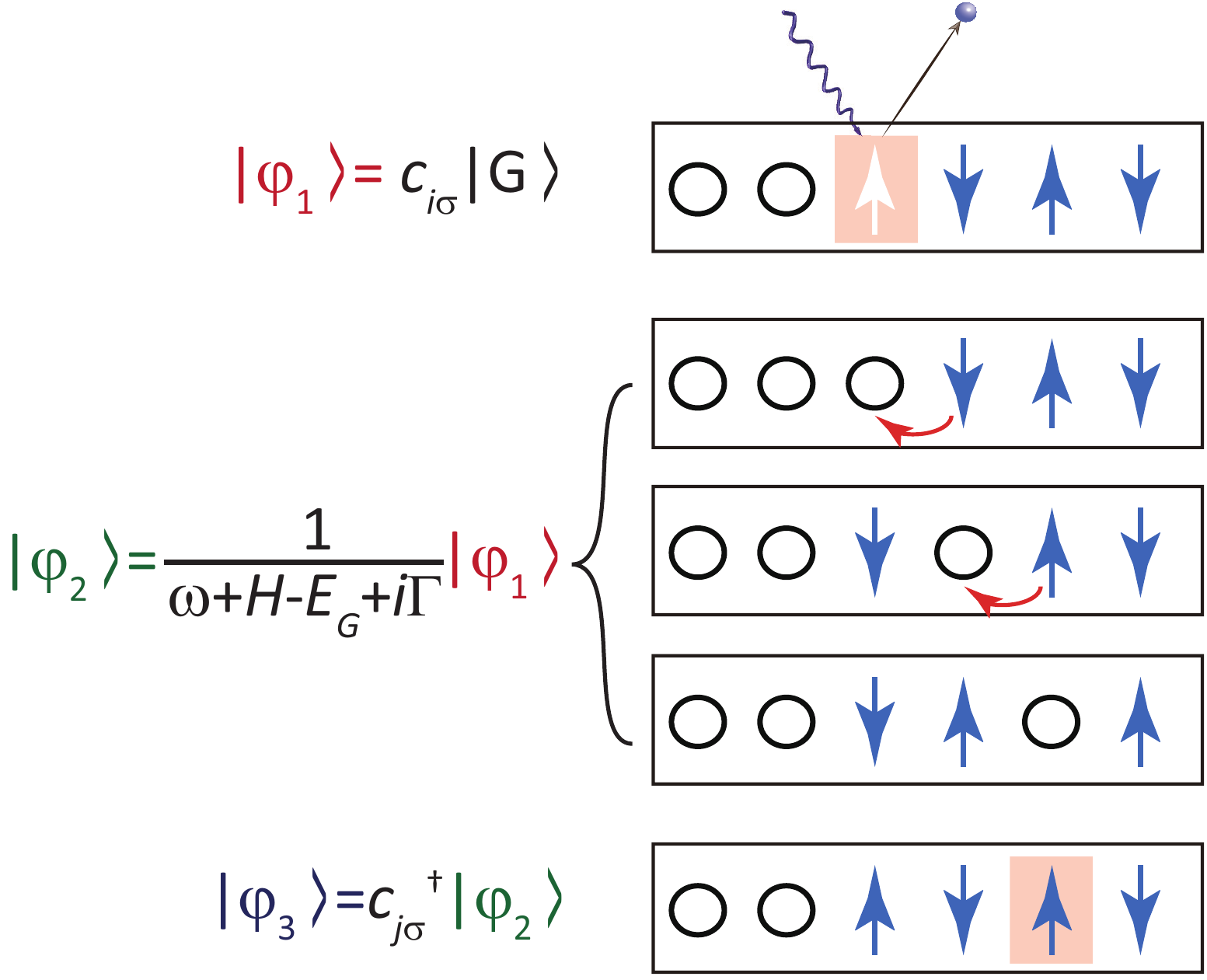}
\caption{\label{fig:cartoonb} Cartoons illustrating another channel for the coherent motion of a photo-doped hole in the $t$--$J$ model with a finite density of holes in the ground state.
This channel contributes only when $J \neq 0$.
The hole can move coherently as the spin exchange process $\propto J$ may repair the `defects' made by the hole motion, so that a `kinetically relaxed spin polaron' can form.
The photoemission spectrum is related to the imaginary part of the Green's function $\langle \varphi_3|\varphi_1\rangle$.
For simplicity, the cartoons only show hole motion along a 1D path in a 2D lattice.}
\end{center}
\end{figure}

Altogether, the dependence of the spectral functions on the spin exchange $J$ can be rationalized in the following way.
Firstly, the inverse photoemission spectrum becomes {\it less} coherent for larger $J$, since the spins of the doped photo-electrons can separate from the added charge and lead to a substantial incoherent response. 
Next, for the photoemission spectrum we encounter a situation somewhat similar to the one 
observed for the extremely well-studied case of a single hole in the undoped antiferromagnet. Indeed, as $J$ increases the spectrum becomes more coherent, suggesting that the doped hole can move in a more coherent way by coupling to the spin fluctuations (see the cartoon picture in Fig.~\ref{fig:cartoonb}). 
Naturally, the latter case is far less extreme than the well-known undoped one, as in the latter situation the whole spectrum at $J=0$ is almost completely incoherent whereas in the low hole doping regime for $J=0$ the spectral function is ``semi-coherent'' (see previous subsection). For the same reason, unlike in the undoped case, the bandwidth is reduced with increasing $J$, since already at $J=0$ the hole seems to be able to move in a ``semi-coherent'' way.  

\section{Conclusions}

We have studied the evolution of spectral properties of the hole-doped Mott insulators by applying the CPT method to the Hubbard and $t$--$J$-like models. 
It turns out that for all studied doping levels, which range from 12.5\% to 37.5\% hole doping, the most apparent feature below the Mott gap is a {\it quasi-free dispersion}: except for the spectral weight located well below the Fermi level and close to the $\Gamma$ point, the dominant spectral feature below the Mott gap follows a tight-binding model dispersion relation with a renormalized hopping amplitude. 

The main result of this work concerns understanding the nature of the quasi-free dispersion feature.
We have shown that the observed quasi-free dispersion is not as closely related to the free electron hopping as one would naively judge from its appearance in the Hubbard spectra,
and that the onset of ``gaps'' or ``waterfalls'' in the spectral function of the Hubbard model can be naturally expected.
We suggest that the latter ones originate from strong electron-electron correlations present in the system,
in agreement with the recent study of lightly doped Mott insulating iridium oxide~\cite{Battisti2017}.
This is due to our numerical investigation which shows that both electron-electron correlation and spin exchange interaction {\it inherently} influence the most intensive spectral feature of the Hubbard model:

First of all, we have shown that to understand the nature of the quasi-free dispersion, it is enough to focus on the \tJ\ model with much fewer degrees of freedom.
Rather surprisingly, at the $J=0$ limit in this intrinsically correlated model, the spectral function is far less coherent than that with a canonical value of $J=0.5t$.
In this extremely correlated regime ($U \rightarrow \infty$ once $J \rightarrow 0$ and hopping is finite), the most intensive peaks only very roughly follow the weakly renormalized (\textit{e.g.} by a factor of 0.8 at 12.5\% doping) free dispersion, as they do not form a continuous band in momentum space.
We have provided an intuitive explanation of this ``semi-coherence'' by suggesting circumstances when the coherent motion of added charge carriers can occur in the $t$--$J$ model with $J=0$, but this should be regarded as an exception rather than a rule. 
This means that the studied case lies somewhat in between the known examples of completely incoherent motion of a single hole in the {\it undoped} $t$--$J$ model with $J=0$~\cite{Brinkman1970}, and the completely coherent motion of holes predicted in a spinless fermion or hard-core boson model. 
We also note that the simple Gutzwiller mean-field picture~\cite{Zhang1988} fails to predict the observed band renormalization. 

Next, we have studied the spectral function of the \tJ\ model with finite spin exchange interaction $J$ and finite doping.
The increase of the spin exchange $J$ leads to a more continuous dispersion relation formed by the most intensive peaks in the spectrum.
At the same time, the impact of spin exchange $J$ on the coherence of single-particle excitations (observed in the spectrum as the broadening of the most intensive peaks in the spectral function) is opposite below and above the Fermi level.
Due to the coupling of moving holes to spin fluctuations -- a situation somewhat similar to the case of a single hole in an undoped antiferromagnetic ground state -- 
the spectra below the Fermi level become more coherent with increasing $J$.
On the other hand, inverse photoemission spectrum displays a more incoherent response with increasing $J$, possibly due to the fact that 
the spins of photoinduced electrons above the Fermi level can separate from the added charge. 
In both cases, however, the coupling to spin excitations makes the photoinduced carriers less mobile.
As a result, the bandwidth of the quasi-free dispersion is more strongly renormalised for a finite $J$ than for the $J=0$ case.

\section*{Acknowledgments}
We thank  A.~Kanigel, M.~Ma\'ska, A.~Moreo, A.~M.~Ole\'s, Z.-X.~Shen, J.~Spa\l{}ek, and M.~Wysoki\'nski, for insightful discussions. This work was supported at SLAC and Stanford University by the U.S. Department of Energy, Office of Basic Energy Sciences, Division of Materials Sciences and Engineering, under Contract No. DE-AC02-76SF00515. Y.~W. was supported by the Stanford Graduate Fellows in Science and Engineering. A portion of the computational work was performed using the resources of the National Energy Research Scientific Computing Center supported by the U.S. Department of Energy, Office of Science, under Contract No.~DE-AC02-05CH11231.
K.~W. acknowledges support from Narodowe Centrum Nauki (NCN), under Project No.~2016/22/E/ST3/00560 and Project No.~2016/23/B/ST3/00839 .

 \bibliographystyle{prsty}

\begin{thebibliography}{10}

\bibitem{RevModPhys.70.1039}
M. Imada, A. Fujimori, and Y. Tokura, Rev. Mod. Phys. {\bf 70},  1039  (1998).

\bibitem{lee2006doping}
P.~A. Lee, N. Nagaosa, and X.-G. Wen, Rev. Mod. Phys. {\bf 78},  17  (2006).

\bibitem{Auerbach1994}
A. Auerbach, {\em Interacting Electrons and Quantum Magnetism} (Springer, New
  York, 1994).

\bibitem{Schmitt:1988SCBAThry}
S. Schmitt-Rink, C.~M. Varma, and A.~E. Ruckenstein, Phys. Rev. Lett. {\bf 60},
   2793  (1988).

\bibitem{Kane:1989SCBAThry}
C.~L. Kane, P.~A. Lee, and N. Read, Phys. Rev. B {\bf 39},  6880  (1989).

\bibitem{Martinez:1991SpinPolaron}
G. Martinez and P. Horsch, Phys. Rev. B {\bf 44},  317  (1991).

\bibitem{Efstratios:2007SCBA}
E. Manousakis, Phys. Rev. B {\bf 75},  035106  (2007).

\bibitem{Zemljic:2008}
M.~M. Zemlji\ifmmode~\check{c}\else \v{c}\fi{}, P.
  Prelov\ifmmode~\check{s}\else \v{s}\fi{}ek, and T. Tohyama, Phys. Rev. Lett.
  {\bf 100},  036402  (2008).

\bibitem{Preuss:1995SDW}
R. Preuss, W. Hanke, and W. von~der Linden, Phys. Rev. Lett. {\bf 75},  1344
  (1995).

\bibitem{macridin2007high}
A. Macridin, M. Jarrell, T. Maier, and D. Scalapino, Phys. Rev. Lett. {\bf 99},
   237001  (2007).

\bibitem{stephan1992single}
W. Stephan and P. Horsch, Int. J Mod. Phys. B {\bf 6},  589  (1992).

\bibitem{tJ3s}
J.~H. Jefferson, H. Eskes, and L.~F. Feiner, Phys. Rev. B {\bf 45},  7959
  (1992).

\bibitem{Bala:1995ttprJSCBA}
J. Ba{\l}a, A. Ole{\'s}, and J. Zaanen, Phys. Rev. B {\bf 52},  4597  (1995).

\bibitem{wang2015origin}
Y. Wang {\it et~al.}, Phys. Rev. B {\bf 92},  075119  (2015).

\bibitem{vojta2009lattice}
M. Vojta, Advances in Physics {\bf 58},  699  (2009).

\bibitem{LeBlanc2015}
J.~P.~F. LeBlanc {\it et~al.}, Phys. Rev. X {\bf 5},  041041  (2015).

\bibitem{jia2014persistent}
C. Jia {\it et~al.}, Nat. Phys. {\bf 5},  3314  (2014).

\bibitem{Kung2015}
Y.~F. Kung {\it et~al.}, Phys. Rev. B {\bf 92},  195108  (2015).

\bibitem{huang2016numerical}
E.~W. Huang {\it et~al.}, arXiv:1612.05211  (2016).

\bibitem{TKLee2016}
W.-J. Li, C.-J. Lin, and T.-K. Lee, Phys. Rev. B {\bf 94},  075127  (2016).

\bibitem{Kung2017}
Y.~F. Kung {\it et~al.}, Phys. Rev. B {\bf 96},  195106  (2017).

\bibitem{Braicovich2010}
L. Braicovich {\it et~al.}, Phys. Rev. B {\bf 81},  174533  (2010).

\bibitem{le2011intense}
M. Le~Tacon {\it et~al.}, Nat. Phys. {\bf 7},  725  (2011).

\bibitem{dean2013persistence}
M. Dean {\it et~al.}, Nat. Mater. {\bf 12},  1019  (2013).

\bibitem{Dean_PRL_2013}
M. Dean {\it et~al.}, Phys. Rev. Lett. {\bf 110},  147001  (2013).

\bibitem{lee2014asymmetry}
W. Lee {\it et~al.}, Nat. Phys. {\bf 10},  883  (2014).

\bibitem{Guarise2014}
M. {Guarise} {\it et~al.}, Nature Communications {\bf 5},  5760  (2014).

\bibitem{ishii2014high}
K. Ishii {\it et~al.}, Nat. Commun. {\bf 5},  3714  (2014).

\bibitem{Huang2016}
H.~Y. {Huang} {\it et~al.}, Scientific Reports {\bf 6},  19657  (2016).

\bibitem{Minola2017}
M. Minola {\it et~al.}, Phys. Rev. Lett. {\bf 119},  097001  (2017).

\bibitem{tranquada2004quantum}
J. Tranquada {\it et~al.}, Nature {\bf 429},  534  (2004).

\bibitem{tranquada2007neutron}
J.~M. Tranquada,  in {\em Handbook of High-Temperature Superconductivity},
  edited by J.~R. Schrieffer and J.~S. Brooks (Springer, New York, 2007),
  Chap.~6.

\bibitem{Stephan1991}
W. Stephan and P. Horsch, Phys. Rev. Lett. {\bf 66},  2258  (1991).

\bibitem{Moreo1995}
A. Moreo, S. Haas, A.~W. Sandvik, and E. Dagotto, Phys. Rev. B {\bf 51},  12045
   (1995).

\bibitem{Haas1995}
S. Haas, A. Moreo, and E. Dagotto, Phys. Rev. Lett. {\bf 74},  4281  (1995).

\bibitem{Wen1996}
X.-G. Wen and P.~A. Lee, Phys. Rev. Lett. {\bf 76},  503  (1996).

\bibitem{Duffy1997}
D. Duffy {\it et~al.}, Phys. Rev. B {\bf 56},  5597  (1997).

\bibitem{Byczuk2007}
K. {Byczuk} {\it et~al.}, Nature Physics {\bf 3},  168  (2007).

\bibitem{Moritz:2009fb}
B. Moritz {\it et~al.}, New J Phys. {\bf 11},  093020  (2009).

\bibitem{kohno2015spectral}
M. Kohno, Phys. Rev. B {\bf 92},  085128  (2015).

\bibitem{Sakai2016L}
S. Sakai, M. Civelli, and M. Imada, Phys. Rev. Lett. {\bf 116},  057003
  (2016).

\bibitem{Sakai2016}
S. Sakai, M. Civelli, and M. Imada, Phys. Rev. B {\bf 94},  115130  (2016).

\bibitem{shastry2017}
B.~S. Shastry and P. Mai, arXiv:1703.08142  (2017).

\bibitem{Senechal:2000fg}
D. S{\'e}n{\'e}chal, D. Perez, and M. Pioro-Ladri{\`e}re, Phys. Rev. Lett. {\bf
  84},  522  (2000).

\bibitem{Maska1998}
M.~M. Ma\ifmmode~\acute{s}\else \'{s}\fi{}ka, Phys. Rev. B {\bf 57},  8755
  (1998).

\bibitem{Senechal:2002fr}
D. S{\'e}n{\'e}chal, D. Perez, and D. Plouffe, Phys. Rev. B {\bf 66},  075129
  (2002).

\bibitem{aichhorn2005weak}
M. Aichhorn and E. Arrigoni, Rev. Mod. Phys. {\bf 72},  117  (2005).

\bibitem{senechal2012book}
D. S{\'e}n{\'e}chal,  in {\em Strongly Correlated Systems}, edited by A. Avella
  and F. Mancini (Springer, Berlin - Heidelberg, 2012), pp.\ 237--270.

\bibitem{pairault1998strong}
S. Pairault, D. S{\'e}n{\'e}chal, and A.-M. Tremblay, Phys. Rev. Lett. {\bf
  80},  5389  (1998).

\bibitem{Kohno:2012PRL}
M. Kohno, Phys. Rev. Lett. {\bf 108},  076401  (2012).

\bibitem{PhysRevB.90.035111}
M. Kohno, Phys. Rev. B {\bf 90},  035111  (2014).

\bibitem{anderson1987resonating}
P.~W. Anderson, Science {\bf 235},  1196  (1987).

\bibitem{Zhang:1988jf}
F. Zhang and T. Rice, Phys.Rev. B {\bf 37},  3759  (1988).

\bibitem{Eskes:1988ef}
H. Eskes and G. Sawatzky, Phys. Rev. Lett. {\bf 61},  1415  (1988).

\bibitem{PhysRevB.44.10256}
Q. Li, J. Callaway, and L. Tan, Phys. Rev. B {\bf 44},  10256  (1991).

\bibitem{Dagotto1992}
E. Dagotto, F. Ortolani, and D. Scalapino, Phys. Rev. B {\bf 46},  3183
  (1992).

\bibitem{PhysRevB.47.1160}
M. Veki\ifmmode~\acute{c}\else \'{c}\fi{} and S. White, Phys. Rev. B {\bf 47},
  1160  (1993).

\bibitem{Groeber2000}
C. Gr\"ober, R. Eder, and W. Hanke, Phys. Rev. B {\bf 62},  4336  (2000).

\bibitem{becca_sorella_2017}
F. Becca and S. Sorella, {\em Quantum Monte Carlo Approaches for Correlated
  Systems} (Cambridge University Press, Cambridge, 2017).

\bibitem{Georges1996}
A. Georges, G. Kotliar, W. Krauth, and M.~J. Rozenberg, Rev. Mod. Phys. {\bf
  68},  13  (1996).

\bibitem{kancharla2008anomalous}
S. Kancharla {\it et~al.}, Phys. Rev. B {\bf 77},  184516  (2008).

\bibitem{weber2010strength}
C. Weber, K. Haule, and G. Kotliar, Nat. Phys. {\bf 6},  574  (2010).

\bibitem{sakai2009evolution}
S. Sakai, Y. Motome, and M. Imada, Phys. Rev. Lett. {\bf 102},  056404  (2009).

\bibitem{sakai2010doped}
S. Sakai, Y. Motome, and M. Imada, Physical Review B {\bf 82},  134505  (2010).

\bibitem{Schollwoeck2005}
U. Schollw\"ock, Rev. Mod. Phys. {\bf 77},  259  (2005).

\bibitem{yuan2005doping}
Q. Yuan, F. Yuan, and C. Ting, Phys. Rev. B {\bf 72},  054504  (2005).

\bibitem{arrigoni2009phase}
E. Arrigoni, M. Aichhorn, M. Daghofer, and W. Hanke, New Journal of Physics
  {\bf 11},  055066  (2009).

\bibitem{gull2013superconductivity}
E. Gull, O. Parcollet, and A.~J. Millis, Phys. Rev. Lett. {\bf 110},  216405
  (2013).

\bibitem{han2016charge}
X.-J. Han {\it et~al.}, New Journal of Physics {\bf 18},  103004  (2016).

\bibitem{Potthoff_PRB_2003}
M. Potthoff, M. Aichhorn, and C. Dahnken, Phys. Rev. Lett. {\bf 91},  206402
  (2003).

\bibitem{pairault2000strong}
S. Pairault, D. Senechal, and A.-M. Tremblay, The European Physical Journal
  B-Condensed Matter and Complex Systems {\bf 16},  85  (2000).

\bibitem{Zhang1988}
F.~C. Zhang, C. Gros, T.~M. Rice, and H. Shiba, Superconductor Science and
  Technology {\bf 1},  36  (1988).

\bibitem{chao1977tJmodel}
K. Chao, J. Spalek, and A. Ole\'{s}, J Phys C: Solid State Phys {\bf 10},  L271
   (1977).

\bibitem{chao1978canonical}
K. Chao, J. Spa{\l}ek, and A. Ole{\'s}, Phys. Rev. B {\bf 18},  3453  (1978).

\bibitem{belinicher1994consistent}
V. Belinicher and A. Chernyshev, Phys. Rev. B {\bf 49},  9746  (1994).

\bibitem{belinicher1994range}
V. Belinicher, A. Chernyshev, and L. Popovich, Lett. B {\bf 50},  13768
  (1994).

\bibitem{Spalek1988}
J. Spa\l{}ek, Phys. Rev. B {\bf 37},  533  (1988).

\bibitem{Szczepanski1990}
K.~J. von Szczepanski, P. Horsch, W. Stephan, and M. Ziegler, Phys. Rev. B {\bf
  41},  2017  (1990).

\bibitem{Eskes1994}
H. Eskes and A.~M. Ole\ifmmode~\acute{s}\else \'{s}\fi{}, Phys. Rev. Lett. {\bf
  73},  1279  (1994).

\bibitem{psaltakis1992optical}
G.~C. Psaltakis, Phys. Rev. B {\bf 45},  539  (1992).

\bibitem{Brinkman1970}
W.~F. Brinkman and T.~M. Rice, Phys. Rev. B {\bf 2},  1324  (1970).

\bibitem{Ruckenstein1988}
A.~E. Ruckenstein and S. Schmitt-Rink, Phys. Rev. B {\bf 38},  7188  (1988).

\bibitem{Fang1989}
Y. Fang, A.~E. Ruckenstein, E. Dagotto, and S. Schmitt-Rink, Phys. Rev. B {\bf
  40},  7406  (1989).

\bibitem{Gros1989}
C. Gros and M.~D. Johnson, Phys. Rev. B {\bf 40},  9423  (1989).

\bibitem{Gehlhoff1995}
L. Gehlhoff and R. Zeyher, Phys. Rev. B {\bf 52},  4635  (1995).

\bibitem{Gurin2001}
P. Gurin and Z. Gulacsi, Philosophical Magazine B {\bf 81},  321  (2001).

\bibitem{Foussats2002}
A. Foussats and A. Greco, Phys. Rev. B {\bf 65},  195107  (2002).

\bibitem{Bejas2006}
M. Bejas, A. Greco, and A. Foussats, Phys. Rev. B {\bf 73},  245104  (2006).

\bibitem{Footnote3}
This was also confirmed by the ED calculations (unshown), i.e. it is not an
  artefact of the CPT method.

\bibitem{TrugmanPath}
S.~A. Trugman, Phys. Rev. B {\bf 37},  1597  (1988).

\bibitem{Vojta1998}
M. Vojta and K.~W. Becker, Phys. Rev. B {\bf 57},  3099  (1998).

\bibitem{Footnote4}
In fact, in our numerical calculations we have assumed that the number of spin
  up and spin down electrons is the same. This way the constrained fermion
  model more closely resembles the $t$--$J$ model with a finite $J$. However,
  if this condition were relaxed, we would find even stronger and longer range
  ferromagnetic correlations.

\bibitem{Battisti2017}
I. {Battisti} {\it et~al.}, Nature Physics {\bf 13},  21  (2017).

\end{thebibliography}

\end{document}